\documentclass[12pt,a4paper]{article}
\usepackage{jstyle}
\usepackage{slashed}

\newcommand{\bi}{\begin{itemize}}
\newcommand{\ei}{\end{itemize}}
\newcommand{\rmd} {{\rm d}}

\newcommand{\bea}{\begin{eqnarray}}
\newcommand{\eea}[1]{\label{#1}\end{eqnarray}}
\newcommand{\beq}{\begin{equation}}
 \newcommand{\eeq}[1]{\label{#1}\end{equation}}

	\def\calJ{\mathcal{J}}
	\def\calK{\mathcal{K}}
	\def\calL{\mathcal{L}}
	\def\frb{\mathfrak{b}}

	\def\fsl2{\mathfrak{sl}(2 ,\mathbb R)}

	\def\a{\alpha}
	\def\b{\beta}

	\def\d{\delta}
	\def\r{\rho}
	
	\def\l{\lambda}
	\def\e{\epsilon}
	
	\def\m{\mu}
	\def\n{\nu}
	\def\k{\kappa}
	
	\def\vph{\varphi}
	
	\def\({\left(}
	\def\){\right)}
	\def\[{\left[}
	\def\]{\right]}

\author[a]{Euihun JOUNG}
\author[b,c]{\quad Jaewon KIM}
\author[d]{\quad Jihun KIM}
\author[b]{\quad Soo-Jong REY}

\affiliation[a]{Department of Physics and Research Institute of Basic Science \\ Kyung Hee University, Seoul 02447 \rm KOREA
}
\affiliation[b]{School of Physics \& Astronomy, Seoul National University, Seoul 08826 \rm KOREA}
\affiliation[c]{Department of Emerging Materials Science, DGIST, Daegu 42988 \rm KOREA}
\affiliation[d]{Department of Physics and Center for Cosmology \& Particle Physics \\
New York University, New York 10003 \rm USA}

\emailAdd{euihun.joung@khu.ac.kr, 
fgwanabe@gmail.com}
\emailAdd{
jk2943@nyu.edu, 
rey.soojong@gmail.com}

\title{\centering
\LARGE{
Asymptotic Symmetries of \\
Colored Gravity in Three Dimensions 
}}

\abstract{
Three-dimensional colored gravity refers to nonabelian isospin extension of Einstein gravity. We investigate the asymptotic symmetry algebra of  the $SU(N)$-colored gravity in (2+1)-dimensional anti-de Sitter spacetime. Formulated by the Chern-Simons theory with $SU(N,N)\times SU(N,N)$ gauge group,  the theory contains graviton, $SU(N)$ Chern-Simons gauge fields and massless spin-two multiplets in the $SU(N)$ adjoint representation, thus extending diffeomorphism to colored, nonabelian counterpart.
We identify the asymptotic symmetry as Poisson algebra of generators associated with the residual global symmetries of the nonabelian diffeomorphism set by  appropriately chosen boundary conditions. The resulting asymptotic symmetry algebra 
is a nonlinear extension of 
Virasoro algebra and $\widehat{\mathfrak{su}(N)}$ Kac-Moody algebra, supplemented by additional generators corresponding to the massless spin-two adjoint matter fields.
}

\begin{document}


\maketitle

\rightline{\sl The Sun also rises,}
\rightline{-- Ernest  \ Hemingway}
\rightline{\sl again, again, and again.}

\section{Introduction}

Einstein gravity in three-dimensional anti-de Sitter (adS) spacetime sits at the crossroad of the simplicity of lower-dimensional spacetime and the richness of higher-dimensional spacetime. There is no propagating massless spinning fields but there are non-trivial Banados-Teitelboim-Zanelli (BTZ) black holes, be they neutral, charged or rotating \cite{BTZ}.
In three-dimensional spacetime, the Einstein gravity also accommodates many interesting extensions including, notably, the higher-spin gravity \cite{Blencowe, PV}. 

Recently, we developed a new extension of the three-dimensional Einstein gravity in \cite{Col G, Col HS}. The extension can be summarized most transparently in the Chern-Simons (CS) formulation. At the classical level, the pure Einstein gravity on AdS$_3$ is equivalent to the CS theory with gauge group $SL(2,\mathbb R)_L \times SL(2,\mathbb R)_R$ \cite{AT, W3}. The idea is that we decorate each copy of the gauge group $SL(2,\mathbb R)$ with a `color'  isospin symmetry group, for example, $U(N)$.
Naively, this would mean that we replace the usual $\mathfrak{sl}(2,\mathbb{R})$ valued gauge connection with 
the $\mathfrak{sl}(2, \mathbb{R}) \otimes \mathfrak{u}(N)$ valued ones.
Although the no-go theorem \cite{nogo}\footnote{See \cite{Nico} for an ``exotic'' exception.}  prevents the tensor-product vector space $\mathfrak{sl}(2,\mathbb{R})\otimes \mathfrak{u}(N)$
to admit a Lie algebra structure, we can get around it by enlarging the vector space with additional $\mathfrak{su}(N)$
to obtain a vector space furnished with a Lie algebra isomorphic to $\mathfrak{su}(N,N)$:
\beq
\mathfrak{sl}(2,\mathbb{R})\otimes \mathfrak{u}(N)
\oplus \mathfrak{su}(N)
 \simeq \mathfrak{su}(N,N)\,.
\eeq{al} 
The upshot of this enlargement is the $SU(N,N)$ CS theory describing not only multiple of massless spin-two fields,
but also the $SU(N)_L\times SU(N)_R$ CS gauge fields.
The $1+ (N^2-1)$ copies of massless spin-two fields
consist of the genuine graviton and  the rest whose
left and right moving modes carry the adjoint representations of the $SU(N)_L$  and $SU(N)_R$, respectively.
We call the corresponding gravity theory $SU(N)$ colored gravity theory.

The $SU(N)$ colored gravity theory has many interesting features. 
In particular, in contrast to uncolored counterpart, the colored theory admits  multiple vacua, referred to as \emph{rainbow vacua} in \cite{Col G}, 
breaking the $SU(N)$ color isospin symmetry as
\beq
SU(N) \quad \rightarrow  
\quad G=SU(N-k)\times SU(k)\times U(1)\,.
\eeq{broken}
The \emph{breaking parameter} $k$ ranges over between $0$ and $\left[ \tfrac{N+1}{2} \right]$,
and each of these vacua corresponds to an AdS background with a $k$-dependent curvature scale. 
The $k=0$ corresponds to the color-singlet vacuum, where the theory has the perturbative spectrum described above:
{\it i)} graviton, {\it ii)} $SU(N)\times SU(N)$ CS gauge fields, and {\it iii)} $SU(N)$ adjoint massless spin-two fields.
In the vacuum with non-zero $k$, due to the symmetry breaking~(\ref{broken}), the theory has different perturbative degrees of freedom. It has {\it i)}  graviton,
{\it ii)}  $G\times G$ CS  gauge fields, 
{\it iii)} $G$ adjoint massless spin-two fields,
and {\it iv)} the fields associated with the $2\,(N-k)\,k$ broken generators in the coset $SU(N)/G$. In this symmetry-broken sector,
a Higgs-like mechanism glues the massless spin-two fields with
the spin-one, namely, the left and right CS fields.
The resulting spectrum turns out to be partially-massless spin-two fields \cite{PM,PM'}, carrying now the bi-fundamental representation of the residual symmetry $G$. If the background were de Sitter (dS) spacetime, partially-massless spin-two field corresponds to a massive spin-two field whose mass lies exactly on the unitarity bound known as the Higuchi bound \cite{HB}. 
On this bound, a scalar-parameter gauge symmetry persists and the helicity zero mode decouples leaving only the helicity two and one modes.
If the background were AdS spacetime, the same mechanism holds except that the bound is not the unitarity one but the point where the norm of tachyonic spin-two field changes the sign of the helicity-zero mode.
Still, the helicity-one modes have negative norms with respect to the helicity-two modes, and so the partially-massless fields in AdS background are non-unitary. Despite of the non-unitarity, it plays many interesting roles in a variety of contexts  (see e.g. \cite{PM1, PM2}). Interestingly, the partially-massless fields persist to exist in three dimensions, despite of the fact that all the helicity modes are not propagating \cite{Col G, Col HS}.

By the AdS/CFT correspondence, there must be two-dimensional conformal field theories (CFTs) dual to the panorama of colored gravity. Spacetime and internal symmetries of these CFTs are deducible from asymptotic symmetry algebra of each rainbow vacua of the colored gravity.  As is well-known, slow fall-off of gravitational field in three-dimensional AdS spacetime enhances the asymptotic symmetry algebra to infinite-dimensional affine extensions. For the colored gravity at hand, the diffeomorphism is color-decorated, so the asymptotic symmetry algebra dictated by large colored diffeomorphism would be much richer. Roughly speaking, we anticipate color-decorated Brown-Henneaux algebra. 

As said, the rainbow vacua and spectrum therein point to panorama of the asymptotic symmetries. To study these symmetries systematically, one would first construct the CFT dual to colored gravity in the $k=0$ color-singlet vacuum and, from this, derive the CFTs dual to $k \ne 0$ color-broken vacua. Along the way, one ought to face the issue of unitarity. In the color-broken vacua, the spectrum includes partially-massless fields, 
{whose boundary degrees of freedom carry non-unitary representations of the AdS isometry, $so(2,2)$
(see the section 6 of \cite{Col HS}).}
Moreover, among the $SU(N-k)\times SU(k)\times U(1)$ adjoint massless spin-two fields, the ones corresponding to the $SU(k)\times U(1)$  turn out to be ghosts as  they have the negative sign of kinetic term with respect to the sign of graviton kinetic term. Intuitively, one can understand  
{the sign flip of kinetic terms} from the shape of effective potential of the colored spin-two matters from which the rainbow vacua have been derived.  In fact, the rainbow vacuum solutions are all saddle points except for the $k=0$ color-singlet one. {All these seem to suggest that the colored gravity in a $k\neq0$ background has an issue of unitarity. Since $k$ is a parameter that characterizes the pattern of spontaneous gauge symmetry breaking and so the pattern of mass spectrum reorganizations, it is hard to imagine that the ``amount'' of non-unitarity depends on the value of $k$. Therefore, we are naturally lead to the suspicion that, if the background with $k \ne 0$ has an issue of unitarity, the color-singlet background at $k=0$ ought to have the same issue in a hidden manner.
Around the color-singlet background at $k=0$, all the massless spin-two fields including the graviton are unitary, so the only source one would suspect is the $SU(N)\times SU(N)$ CS part. Usually, one does not question the unitarity of CS theory as it is a matter of definition in so far as gauge invariance is maintained. Here, however, we do. }

With the aim to find answers to the above two questions, in this paper, we study the asymptotic symmetry of the $SU(N)$ colored gravity with particular attention to pin down the holography of non-unitarities. The asymptotic symmetry of three-dimensional gravity, and its cousins, has played a central role in the development of string theory black holes and holography since the seminal work by Brown and Henneaux \cite{BH}. Recent advances extended the scope to the asymptotic symmetry of higher spin gravity \cite{CFPT,AS}, which led to a version of AdS$_3$/CFT$_2$ proposal \cite{GaGo}. {Here, we explicitly derive an asymptotic symmetry algebra of the $SU(N)$ colored gravity and discover that the $SU(N) \times SU(N)$ CS part of the theory indeed contains the non-unitarity: the boundary excitation modes of $SU(N) \times SU(N)$  spin-one dynamics 
have negative norm with respect to the gravity modes.}

We organized this paper as follows.
In Section \ref{sec: 2}, we recapitulate salient features of the colored gravity on AdS$_3$ relevant for the foregoing considerations. In Section \ref{sec: 3}, we impose asymptotically AdS$_3$ boundary conditions on colored gravity and obtain gauge transformations preserving them. We observe that resulting gauge transformations and variations of fields have non-linear dependence on fields. In Section \ref{sec: 4}, we read off Poisson brackets between dynamical variables. The algebra is non-linear in terms of fields which do not transform as Virasoro primaries. We show that field redefinition of Virasoro generator make all fields into Virasoro primaries but the non-linearity is not removable. In Section \ref{sec: 5}, we discuss negative level of the affine algebra and its relation to the non-unitarity.  Appendix \ref{app} contains additional technical details.

\section{Colored Gravity}\label{sec: 2}

Before starting the analysis of asymptotic symmetries of colored gravity,
we first recapitulate salient features of the colored gravity.

The $SU(N)$ color-decorated gravity in three dimensions can be described by the CS action with the gauge algebra $\mathfrak{su}(N,N)\oplus \mathfrak{su}(N,N)$\,. For the colored-gravity interpretation, we decompose the generators of $\mathfrak{su}(N,N)$ as
\be
	\mathfrak{su}(N,N)\simeq 
	\mathfrak{gl}(2,\mathbb R)\otimes \mathfrak{u}(N)
	\ominus \rm{id}\otimes \mathbb I\,,
	\label{g}
\ee
where the natural associative structures of $\mathfrak{gl}(2,\mathbb R)$
and $\mathfrak{u}(N)$ provide the $\mathfrak{su}(N,N)$  Lie algebra structure in the end:
see Appendix \ref{app} for further details.
The $\mathfrak{gl}(2,\mathbb R)$ is generated by $\{I, J_a\}$ subject to the product,
\be
	J_{a}\,J_{b}=\eta_{ab}\,I+\e_{abc}\,J^{c}\,, 
	\qquad (I,J_a)^\dagger=(I,-J_a)\qquad \qquad (\,a,b,c =0, 1, 2\,)\,,
	\label{gl2}			
\ee
where the $\eta_{ab}$ is the invariant metric with mostly positive signs
and $\epsilon_{abc}$ is the Levi-civita tensor  with sign convention $\epsilon_{012}=+1$\,. 
The $\mathfrak{u}(N)$ algebra is generated by $\{\bm I\,, \bm T_I\}$ with
\be
	\bm{T}_I\,\bm{T}_J=
	\frac{1}{N}\,\delta_{IJ} {\bm I}
	+\left(g_{IJ}{}^{K}+i\,f_{IJ}{}^{K}\right)\bm{T}_K\,, \qquad
	{\bm T_I}^\dagger=\bm T_I\qquad
	(I, J, K = 1, \ldots, N^2 - 1)\,.
	\label{u(N)}
\ee
The totally symmetric and anti-symmetric structure constants $g_{IJK}$ and $f_{IJK}$ are both real-valued. Only identities of $\mathfrak{gl}(2,\mathbb R)$ and $\mathfrak{u}(N)$
admit a non-trivial trace, which are normalized as
\be
	\tr(I)=2\,, 
	\qquad
	\tr(\bm{I})=N\,. 
\ee
By the tensor product structure, the trace in $\mathfrak{su}(N,N)$ is defined as the product of the above traces.

The $SU(N)$ colored gravity is defined by two copies of the CS action,
\be
	S
	=\frac{\k}{4\pi}
	\int \left[\tr\left(\cA\wedge d\cA+\frac23\,\cA\wedge\cA\wedge\cA\right)
	-\tr\left(\tilde\cA\wedge d\tilde\cA+\frac23\,\tilde\cA\wedge\tilde\cA\wedge\tilde\cA\right)\right],
\ee
where the `left-moving' connection one-form can be decomposed into
\be
	\cA=\left(\frac1\ell\,e^{a}+\o^{a}\right) J_{a}\,\bm I
	+\bm A+\frac1{\ell}\,\bm\varphi^{a}\,J_{a}\,.
\ee
Here, $\bm A=A^{I}\,I\,\bm T_{I}$ and $\bm \varphi^{a}=\varphi^{aI}\,\bm T_{I}$
are the gauge field and the colored massless spin-two fields, respectively.  Both of them take
 values in the adjoint representation of $\mathfrak{su}(N)$\,.
They satisfy 
\be
	(\bm A,\bm\varphi^{a})^{\dagger}=(-\bm A,\bm\varphi^{a})\,.
	\label{reality}
\ee
The `right-moving' connection one-form is analogously decomposed with 
$\frac1\ell\,e^{a}-\o^{a}$ and tilded fields as components. 
In \cite{Col G}, it was shown that, by solving the torsion condition $\delta S/\delta \o^a=0$,  the above CS action can be reduced to 
the form
\be
	S
	=S_{\rm CS}
	+\frac{1}{16\pi\,G}\int \rmd^{3}x\sqrt{|g|}\left[
	R-V(\bm\varphi,\tilde{\bm\varphi})+
	\frac{2}{N\,\ell}\,\epsilon^{\m\n\r}\,\tr\left(
	\bm\varphi_{\m}{}^{\l}D_{\n}\bm\varphi_{\r\l}
	-\tilde{\bm\varphi}_{\m}{}^{\l}D_{\n}\tilde{\bm\varphi}_{\r\l}\right)
	\right]. 
\label{final action}
\ee

The above system involves not only the gravity
but also two copies of $SU(N)$ gauge fields
and, most notably, additional massless spin-two matter-like fields transforming as adjoint representation of $SU(N)$\,.
In Eq.\eqref{final action}, the term $S_{\rm CS}$ refers to the two copies of $SU(N)$ CS actions with the level $2\,\k$\,.
The gravitational constant and the AdS radius are related to the level $\k$ as
\be
	\k=\frac{\ell}{4\,N\,G}\,.
	\label{k value}
\ee
The massless spin-two fields are given by the one-derivative action,
but we can also render them to a two-derivative form.
They interact minimally with gravity and $SU(N)$ CS fields through
the covariant derivative,
\be
	D_{\m}\bm\varphi_{\n\r}
	=\nabla_{\m}\bm\varphi_{\n\r}
	+[\bm A_{\mu},\bm\varphi_{\n\r}]\,.
\ee
Moreover, they interact among themselves through the scalar potential  given by
\ba
	&& V(\bm\varphi,\tilde{\bm\varphi}) \nn
	= && -\frac1{N\,\ell^{2}}\,\tr\,\Big[2\,\bm I
	+4\left(\bm\varphi_{[\m}{}^{\m}\,\bm\varphi_{\n]}{}^{\n}
	+\tilde{\bm\varphi}_{[\m}{}^{\m}\,\tilde{\bm\varphi}_{\n]}{}^{\n}
	\right)
	+8
	\left(\bm\varphi_{[\m}{}^{\m}\,\bm\varphi_{\n}{}^{\n}\,\bm\varphi_{\r]}{}^{\r}
	-\tilde{\bm\varphi}_{[\m}{}^{\m}\,\tilde{\bm\varphi}_{\n}{}^{\n}\,
	\tilde{\bm\varphi}_{\r]}{}^{\r}\right)\Big]
	\nn
	&&-\,\frac{16}{N^{2}\,\ell^{2}}\,
	\tr\Big(\bm{\varphi}_{[\m}{}^{\n}\,\bm{\varphi}_{\r]}{}^{\r}-
	\tilde{\bm \varphi}_{[\m}{}^{\n}\,\tilde{\bm \varphi}_{\r]}{}^{\r}\Big)\,
	\tr\Big(\bm{\varphi}_{[\n}{}^{\m}\,\bm{\varphi}_{\l]}{}^{\l}-
	\tilde{\bm \varphi}_{[\n}{}^{\m}\,\tilde{\bm \varphi}_{\l]}{}^{\l}\Big)
	\nn
	&&+\,\frac{6}{N^{2}\,\ell^{2}}\,
	\Big[
	\tr\left(\bm{\varphi}_{[\m}{}^{\m}\,\bm{\varphi}_{\n]}{}^{\n}-
	\tilde{\bm \varphi}_{[\m}{}^{\m}\,\tilde{\bm \varphi}_{\n]}{}^{\n}\right)
	\Big]^2\,.
	\label{V pot}
\ea
Due to the presence of this non-trivial potential, the colored gravity admits a panorama of AdS vacua with different values of the cosmological constant. In this work, we will mainly focus on perturbations around the color-singlet background which is invariant under the $SU(N)$ color symmetry.

\section{Asymptotic Symmetry}\label{sec: 3}

{After its first appearance in analyzing asymptotic dynamics of three-dimensional gravity~\cite{CS AS},
the CS formulation has been applied to compute asymptotic symmetries in more general settings~\cite{CFPT, AS}.
The computing procedure is well-known, 
and we follow it to investigate the asymptotic symmetry of the colored gravity 
around the color-singlet background.
As the `left' and `right' copies of the CS theory are independent from and similar to each other, 
we will only present the analysis for one copy of the CS theory and replicate for the other copy.}

\subsection{Asymptotics for general Chern-Simons theory}

The first step in identifying the asymptotic algebra of colored gravity is in
common with any other gravity theories based on CS formulation.
We briefly recapitulate those steps while keeping the generality, that is, without using specific structures of the colored gravity.

We consider a CS theory with gauge algebra $\mathfrak{g}$ whose base manifold $\cM$ has the topology of $\mathbb R_t\times D_2$\,.
We parameterize the time $\mathbb R_t$ and the disc $D_2$  by
$t$ and the polar coordinates $(\rho,\theta)$, respectively.
We also introduce the light-cone coordinates,
\be
	x^\pm=t\pm \theta\,.
\ee	
The hypersurface $\rho=\infty$ corresponds to the asymptotic boundary with the topology of a cylinder.
In the presence of a boundary, the action must be supplemented by a boundary term or appropriate boundary condition to render the functional variation well-defined.

We primarily consider the left-moving sector \footnote{The analysis for right-moving sector paves the same steps provided the parity inversion $\pm \rightarrow \mp$ is performed at the same time. 
}. One commonly employed boundary condition is $\cA_- \,|_{\partial \cM}=0$\,, that is,
\be
	\lim_{\r\to \infty} \cA_-(\rho,x^+,x^-)=0\,.
	\label{A- condition}
\ee
In the gauge the $\rho$ component of the field is fixed by 
\be
	\cA_{\rho}(\rho,x^+,x^-)= \frb^{-1}(\rho )\,\partial_{\rho}\,\frb(\rho)\,,
	\label{Ar}
\ee
where $\frb(\rho)$ is  {a $G$-valued function (here, $G$ is the Lie group associated to the Lie algebra $\mathfrak{g}$)} that depends
only on the radial coordinate $\rho$\,, the condition \eqref{A- condition} 
along with the flat connection condition $\cF_{\rho - } = 0$ asserts
that the $\cA_-$ component vanishes everywhere:
\be
 	\cA_-(\r,x^+,x^-) = 0\,.
 	\label{A-}
\ee 
The remaining flat connection conditions $\cF_{+ -} =0$ and $\cF_{\rho +}=0$ 
constrain the $\cA_+$ component of the field to the form,
\be
	\cA_+(\r,x^+,x^-) = \frb^{-1}(\rho)\,\mathfrak a(x^+)\,  \frb(\rho)\,,
	\label{A+}
\ee
where the $\mathfrak{a}(x^+)$ is an undetermined function of $x^+$ only.
To summarize, the equations (\ref{Ar}\,--\,\ref{A+})
indicate that the gauge connection $\cA$ is completely specified by two functions $\mathfrak{a}(x^+)$ and $\mathfrak{b}(\rho)$, {taking value in $\mathfrak{g}$ and $G$, respectively.}

The above result was obtained by using all of the flat connection conditions, $\cF=0$, and a part of the gauge symmetry. The rest of the gauge symmetry can still act on fields in the form of gauge function 
\be
	\cC(\r,x^+,x^-)= \frb^{-1}(\rho)\,\mathfrak c(x^+)\,  \frb(\rho)\,.
	\label{L transf}
\ee
From the form of the gauge transformation of $\cA$ under $\cC$\,,
one can find that the Eq.\eqref{L transf} induces the transformation of $\mathfrak{a}$ as
\be
	\delta\,\mathfrak a(x^+)
	=\partial \,\mathfrak c(x^+)+[\mathfrak a(x^+),\mathfrak c(x^+)]\,,
\ee
where $\partial $ means the derivative with respect to $x^+$:
$\partial =\rmd / \rmd x^+$\,.

Thus far, we have relied our analysis on the generic properties of CS theory, but not yet used any specific information of the colored gravity.

\subsection{Asymptotic for Chern-Simons colored gravity}
We now specifically dwell on the CS theory for
the colored gravity with the $SU(N)$ gauge algebra $\mathfrak{g}=\mathfrak{su}(N,N)$, as in Eq.\eqref{g}.
For that, we decompose the $\mathfrak{g}$-valued field $\mathfrak{a}$
as
\be
	\mathfrak{a}(x^+) = u^a(x^+)\,L_a\,\bm I + 
		a^I(x^+)\,I \,\bm T_I + \phi^{aI}(x^+) \,L_a \,\bm T_I \,,
\label{action}
\ee
where we have redefined the generators of $\mathfrak{sl}(2,\mathbb R)$ as
\be
	L_0=-\frac{i}2\,J_0\,,
	\qquad L_\pm=\frac12\left(-i J_1 \pm J_2\right).
\ee
One can understand 
$u^a$, $a^I$, and $\phi^{aI}$ 
as the gravity, the $SU(N)$ CS gauge connection and the colored spin-two matter fields. Note however that their relation to 
the fields  $\frac{e^a}\ell+\o^a,$ $A^I$ and $\varphi^{a,I}$ 
is not diagonal because the adjoint action of $\mathfrak{b}$  mixes different generators. We can similarly decompose the gauge function $\mathfrak{c}$ as
\be
	\mathfrak{c}(x^+) = 
	\x^a(x^+)\,L_a\,\bm I + 
		\l^I(x^+)\,I \,\bm T_I + \z^{aI}(x^+) \,L_a \,\bm T_I \,,
	\label{residual}
\ee
so,  $\x^a$, $\l^I$ and $\z^{aI}$
are associated with the conventional diffeomorphism, the $SU(N)$ gauge symmetry and the nonabelian diffeomorphism (associated with the colored spin-two fields).

We can further restrict the form of fields and residual gauge transformations to 
be compatible with the color-singlet vacuum.  
This means that as we approach the spatial infinity (in other words, in the $\rho\to \infty$ limit) 
 the metric should asymptote to $\rmd s^2 = \ell^2 \left( \rmd \r^2 + e^{2\r}\, \rmd x^+\, \rmd x^- \right)$  while the colored spin-two fields asymptote to $\vph_{\m\n}^I = 0$. 
The $SU(N)$ CS gauge connection is not subject to any condition as 
 {varying it does not alter the asymptotes of the gravity and the colored spin-two fields 
 hence the color-singlet vacuum remains intact.
 Note that the $SU(N)$ part of the connection is similarly unconstrained
 in the analysis of the asymptotic symmetries of extended AdS$_3$ supergravities \cite{Sug}.}

The asymptotic behavior of the metric $g_{\m \n} = e^a{}_\m\,e_{a\n}$ and the colored spin-two fields 
$\vph^I_{\m \n} =e_{a( \m}\, \vph^{aI}{}_{\n )}$ restrict the structure of 
$u^a$ and $\phi^{a,I}$\,.
To proceed further,  we have to fix the function $\frb(\r)$. One particularly convenient choice for the color-single vacuum ($k=0$ case) is
\be
\frb(\r) = \exp \( \r\,L_0\,\bm I\) =  \exp \( \r\,L_0\) \bm I\,.
\label{b}
\ee
This specific form of $\frb(\r)$ and the asymptotic behavior of fields force us to set
\be
u^+ = 1\,, \qquad \phi^{+I} = 0\,.
\label{constraint1}
\ee
{We further simplify the setting with the gauge-fixing condition (see \cite{Balog, Campo} for the issues of the first and second class constraints in this procedure)}
\be
u^0 = 0\,, \qquad \phi^{0I}=0\, .
\label{constraint2}
 \ee
After these, the form of $\mathfrak{a}$ compatible with the colored gravity is 
 \be
	\mathfrak{a}(x^+) = \(L_++u^-(x^+)\,L_-\) \bm I + 
		a^I(x^+)\,I \,\bm T_I + \phi^{-I}(x^+) \,L_- \,\bm T_I \,.
\label{connection}
\ee
So, we have three undetermined functions $u^-(x^+)$, $a^I(x^+)$ and $\phi^{-I}$\,,
 up to the $SU(N)$ color. They correspond to the `left-moving' boundary modes of the graviton, the $SU(N)$ CS gauge connection and the colored spin-two matter field. 
 
\subsection{Asymptotic Symmetry of Colored AdS Gravity} 

Transformation rules of each component of $(u^a,a^I,\phi^{a,I})$ under a gauge transformation generated by $(\x^a,\l^I,\z^{a,I})$ components of $\mathfrak{c}$ gauge function can be calculated using  the relations \eqref{gl2} and \eqref{u(N)}. The results are as follows. 
Firstly, the transformations of the gravity part are given by
\ba
&& \d u^+ = \partial\x^+ + \x^0\,, \\ 
\label{u+}
&& \d u^0 = \partial\x^0 +2\, \x^- -2\,u^-\,\x^+ 
-  \frac{2}{N}\,\phi^{-I}\, \z^{+}{}_{I}\, ,  \\
&& \d u^- = \partial\x^- - u^-\,\x^0 - \frac{1}{N}\,\phi^{-I}\,\z^{0}{}_{I}\,.
\ea
Secondly, the $SU(N)$ CS connection transforms as
\be
\d a^I = \partial\l^{I} + 2\,i\,{f^I}_{JK}\, a^J\, \l^K 
 - i {f^I}_{JK}\,\phi^{-J}\,\z^{+K}\,.
\ee
Finally, the colored spin-two part transforms as
\ba
	&& \d \phi^{+I} = \partial \z^{+I} + \z^{0I} + 
	2\,i\, {f^I}_{JK}\, a^J\, \z^{+K}\,\\
	&& \d \phi^{0I} =\partial \z^{0I}+2\,\z^{-I}
	-2\,u^{-}\,\z^{+I}
	+2\,i\,f^I{}_{JK}\,a^J\,\z^{0K}\nn
	&& \hspace{50pt}
	-\,2\,\phi^{-I}\,\x^{+}
	-2\,g^I{}_{JK}\,\phi^{-J}\,\z^{+K}\, \\
 	&& \d \phi^{-I} =\partial \z^{-I}
 	-u^{-}\,\z^{0I}
 	+2\,i\,f^I{}_{JK}\,a^J\,\z^{-K}\nn
 	&&\hspace{50pt}
 	-\,\phi^{-I}\,\x^{0}
 	+2\,i\,f^I{}_{JK}\,\phi^{-J}\,\l^{K}
 	-g^I{}_{JK}\,\phi^{-J}\,\z^{0K}\,.
 	\label{phi-}
\ea
In the above, we simplified the expressions  (\ref{u+}\,--\,\ref{phi-})
 by using  the asymptotic conditions \eqref{constraint1} and \eqref{constraint2}.
In order to preserve the asymptotic conditions,
the corresponding variation must vanish as well:
\be
	\d u^+=0\,,
	\qquad \d u^0=0\,, 
	\qquad \d \phi^{+I}=0\,,
	\qquad \d\phi^{0I}=0\,.
	\label{asymp cond}
\ee
The asymptotic symmetries of the $SU(N)$ colored gravity
consist of the gauge transformations that satisfy these conditions.
Due to the conditions \eqref{asymp cond},
the asymptotic symmetries are generated only by the parameters,
\be
	\xi:=\xi^+\,,\qquad \l^I\,,\qquad \z^I:=\z^{+I}\,,
\ee
whereas the rest of the parameters are fixed in terms of $(\xi,\l^I,\z^I)$\,.
The gravity part is determined as 
\be
\x^0 = - \partial\x\, , \qquad
\x^- =\frac12\,\partial^2\x  +u^-\,\x+\frac{1}{N}\,\phi^{-I}\, \z_{I}, 
\label{res gauge}
\ee
while the colored spin-two part as
\ba
&& \z^{0I}=-\partial \z^{I} - 
	2\,i\, {f^I}_{JK}\, a^J\, \z^{K}\,,\\
&&\z^{-I}=\frac12
\partial^2 \z^{I}+i\, {f^I}_{JK}\(\partial a^J\, \z^{K}
+2\,a^J\,\partial \z^K\)
	+u^{-}\,\z^{I}\nn
	&&\hspace{50pt}
	+\,2\,f^I{}_{JP}\, {f^P}_{KL}\, a^J\,a^K\, \z^{L}
	+\phi^{-I}\,\x+g^I{}_{JK}\,\phi^{-J}\,\z^{K}\,. 
	\label{lh}
\ea
One can see that the other components of gauge parameter 
depend not only on $(\xi,\l^I,\z^I)$ but also on their derivatives.
Moreover, the parameters involve also a nonlinear dependence
on the field components.

The remaining components of the fields
$(u^-, a^I, \phi^{-I})$ 
correspond to the `left-moving' part of the `would-be' gauge modes of the 
system. They describe
the  boundary degrees of freedom
of the colored gravity.
These components now transform nonlinearly
under the asymptotic symmetries with
parameters $(\xi,\l^I,\z^I)$\,.
Renaming the field components as 
\be
	(u^-,a^I,\phi^{-I})
	 =\frac{2\pi}{\k}\(\frac1N\, \cL\,,\ \cJ^I\, ,\ \cK^I\),
	\label{rename} 
\ee
the gravity part $\cL$ transforms as
\be
\d \calL  = N\frac{\k}{4 \pi} \,\partial^3\,\x
+ \partial\cL\,\x + 2\,\cL\,\partial \x\,
+ \partial\cK^I\,\z_I+ 2\,\cK^I\,\partial \z_I
- i\,\frac{4 \pi}{\k}\, f_{IJK}\,\cJ^I\,\cK^J\,\z^K\,, 
\ee
where the last term exposes a nonlinear dependence on the field.
The $SU(N)$ CS gauge connection part $\cJ^I$ transforms as
\be
\d \cJ^I = \frac{\k}{2 \pi}\,\partial\l^I 
 + 2\, i\, {f^I}_{JK}\,\cJ^J\, \l^K
 -i\, {f^I}_{JK}\,\cK^J\, \z^K\,.
\ee
Finally, the colored spin-two field part $\cK^I$ transforms as
\ba
\d \cK^I \eq
 \frac{\k}{4 \pi} \,\partial^3\z^I
 +\frac{1}{N}\partial\calL\,\z^I  + \frac{2}{N}\,\cL\,\partial\z^I
 +\partial\cK^I\,\x + 2\,\cK^I\,\partial\xi
  \nn
&& +\,i {f^I}_{JK} \( 
\partial^2\cJ^J\,\z^K + 3\,\partial\cJ^J\,\partial \z^K
+ 3\,\cJ^J\,\partial^2\z^K +2 \,\cK^J\,\l^K\) 
+ {g^I}_{JK} \( \partial\cK^J\,\z^K + 2\, \cK^J \,\partial\z^K \)  \nn
&& +\, i \,\frac{4 \pi}{\k}\,{f^I}_{JK} 
\( \cJ^J \,\cK^K\,\x + \frac{2}{N}\,\cL\,\cJ^J \z^K\)  +  i \frac{4 \pi}{\k} 
\({f^I}_{JP}\,{g^P}_{KL} + {g^I}_{KP}\, {f^P}_{JL}\)
\cJ^J\, \cK^K\,\z^L \nn
&& -\,\frac{4 \pi}{\k} \,{f^I}_{JP} \,{f^P}_{KL} 
\(\partial \cJ^J \,\cJ^K\,\z^L + 2\, \cJ^J \,\partial \cJ^K\, \z^L 
+ 3\, \cJ^J\, \cJ^K\,\partial \z^L \)  \nn
&& -\,i\(\frac{4\,\pi}{\k}\)^2
{f^I}_{JQ} {f^Q}_{KR} {f^R}_{LP}\, \cJ^J \,\cJ^K \,\cJ^L\, \z^P\,.
\label{K transf}
\ea
Note that the nonlinearity goes up to the cubic order in the last term.

\section{Non-linear Poisson Algebra from Colored Gravity}\label{sec: 4}
In the previous section, we have derived how the `would-be' gauge modes $(\cL,\cJ^I,\cK^I)$
transform under the asymptotic symmetry.
In order to better understand
the algebraic properties of the symmetry,
we shall identify the Poisson brackets between
the generators of such symmetry.

\subsection{Poisson Bracket}

The variation of the conserved global charge in CS theory is given in general \cite{Charge} by
\be
	\delta Q= -\frac{\k}{2\pi}\int_{\partial \Sigma} \mathrm{d}x^i \, 
	\tr(\cC\, \delta\cA_i)\,.
\ee
In order to identify the charge $Q$ from the above,
we need to `integrate'  the above, taking into account that
the gauge parameter may be field-dependent.
In the case of colored gravity, 
we obtain the charge as
\be
Q =  \int \mathrm{d}\theta \, \left( \x \, \calL  - 2 \lambda_I \, \cJ^I  + \z_I \, \cK^I \right),
\label{Q}
\ee
which correctly reproduces the transformation laws:
 \be
 	\{Q,\calL \} = \delta \calL\,,\qquad
 	\{Q,\cJ^I \} = \delta \cJ^I\,, \qquad 
 	\{Q,\cK^I \} = \delta \cK^I\,.
\ee
Hence, one can simply read off the Poisson bracket between $\calL$, $\cJ^I$ and $\cK^I$ from the expression \eqref{Q} of the charge.

One can first learn that the generator $\cL$ and $\cJ^I$
form a Virasoro and $SU(N)$ Kac-Moody subalgebra:
\ba
& \{ \, \calL (\theta_1), \, \, \calL (\theta_2) \, \} &=\, -\frac{N\,\k}{4\pi}\,\partial^3\delta_{12} - \partial \calL \, \delta_{12} - 2\,\calL \, \partial \delta_{12}\,, \label{PB1} \\
&\{ \cJ^I (\theta_1), \cJ^J (\theta_2) \} &=\,+\frac{\k}{4\pi}\,\delta^{IJ}\, \partial\delta_{12} - i \,f^{IJ}_{\phantom{IJ}K}\,\cJ^K \, \delta_{12}\,, \label{PB2} \\
& \{ \cJ^I (\theta_1), \calL (\theta_2) \} &=\, 0\,. \ \label{PB3}
\ea
The additional generators $\cK^I$ associated with colored spin-two 
transforms covariantly under $\cJ^I$
and $\cL$ up to a nonlinear term for the latter case:
\ba
&\{ \cK^I (\theta_1) , \calL(\theta_2) \} & =\, -\partial\cK^I \, \delta_{12} - 2\cK^I \, \partial \delta_{12} - i\frac{4\pi}{\k}f^{I}_{\phantom{I}JK} \cJ^J \cK^K \delta_{12}\,,\label{PB4} \\
&\{ \cK^I (\theta_1), \cJ^J (\theta_2) \} & =-\, i f^{IJ}_{\phantom{IJ}K} \cK^K \, \delta_{12} \,. \label{PB5}
\ea
The bracket between $\cK^I$'s introduces even higher nonlinearity: 
\ba
&&\{ \cK^I (\theta_1), \cK^J (\theta_2) \} \, =-\, \frac{\k}{4\,\pi}\, \delta^{IJ}\, \partial^3\delta_{12} 
- \frac{\delta^{IJ}}{N} \( \partial\tilde\calL\, \delta_{12} + 2\,\tilde\calL \, \partial \delta_{12} \right) \nn
&&\quad\ +\, i\, f^{IJ}_{\phantom{IJ}K}\left( \partial^2\cJ^K \, \delta_{12} +3\, \partial\cJ^K  \, \partial\delta_{12} +3\, \cJ^K \, \partial^2 \delta_{12} \right) 
- g^{IJ}_{\phantom{IJ}K}\left( \partial\cK^K \, \delta_{12} + 2\,\cK^K \, \partial \delta_{12} \right)\nn
&&\quad\ +\,i\,\frac{8\,\pi}{N\,\k}\, f^{IJ}_{\phantom{IJ}K}\,\calL \, \cJ^K  \, \delta_{12}+i\,\frac{4\,\pi}{\k}\left( f_{K}^{\phantom{K}IN} g_{LN}^{\phantom{LN}J} + g_{L}^{\phantom{L}IN} f_{KN}^{\phantom{KN}J} \right) \cJ^K \cK^L \, \delta_{12} \nn
&&\quad\ +\,\frac{4\,\pi}{\k}\, f_{K}^{\phantom{K}IN} f_{LN}^{\phantom{LN}J} \left(\partial\cJ^K \, \cJ^L \,  \delta_{12} +2\,\cJ^K \, \partial \cJ^L  \, \delta_{12}+3\,\cJ^K  \,\cJ^L \, \partial \delta_{12} \right) \nn
&&\quad\ -\,i \left(\frac{4\,\pi}{\k} \right)^2 \, f_{K}^{\phantom{K}IO} f_{LO}^{\phantom{LO}P} f_{MP}^{\phantom{MP}J}\cJ^K \cJ^L \cJ^M \delta_{12}\,. \label{PB6}
\ea  
In these algebras, all the fields appearing in the right-hand side of the equations are functions of $\theta_1$
and  $\delta_{12}$ is the shorthand notation for $\delta (\theta_1 - \theta_2)$.

For the consistency check of our results, we examined the Jacobi identities of the Poisson brackets (\ref{PB1}\,--\,\ref{PB6}).
The calculation is lengthy, so we just sketch out the procedure. 
In order to make the treatment of multiple delta distributions clearer, 
we introduce the generators with a smearing function $f$\,:
\be 
 \cX[f]=\int d\theta\,\cX(\theta)\,f(\theta)\,,
\ee
and check the Jacobi identity in terms of these `smeared' generators.
Hence, we consider
\begin{align} \label{FoJI}
\{\{ \cX[f_1],\cY[f_2]\},\cZ[f_3]\}
+\{\{ \cY[f_2],\cZ[f_3]\},\cX[f_1]\}
+\{\{ \cZ[f_3],\cX[f_1]\},\cY[f_2]\}\,,
\end{align}
where $\cX$, $\cY$ and $\cZ$ can be any of the generators $\cL$\,, $\cJ^I$ and $\cK^I$\,.
From the double Poisson brackets, we get two delta distributions
whereas we have three integrals coming from each of smeared generators.
In the end,  \eqref{FoJI} will take a form
\begin{align}
\sum_{a,b,c} \int d\theta \, 
\partial^a f_1(\theta)\,\partial^b f_2(\theta)\,\partial^c f_3(\theta)\,\mathcal{F}_{a,b,c}(\theta)\,,
\end{align}
where  $\mathcal{F}$ is a function of $\cL$, $\cJ$ and $\cK$
and their derivatives. 
After removing the ambiguity of integration-by-parts
by setting $c = 0$\,, we have verified $\mathcal{F}_{a,b,0} = 0$ term by term 
(also order by order in $\k$). 
In the latter calculations, various identities of $SU(N)$  structure constants were used:
for instance,
\ba 
&
f_{IJ}^{\phantom{IJ}P} f_{PKQ} + f_{KI}^{\phantom{IK}P} f_{PJQ} + f_{JK}^{\phantom{JK}P} f_{PIQ} = 0\,,\\
&
g_{IJ}^{\phantom{IJ}P} f_{PKQ} + g_{KI}^{\phantom{IK}P} f_{PJQ} + g_{JK}^{\phantom{JK}P} f_{PIQ} = 0\,,
\label{g Jacobi}\\
&
\frac{1}{N}\,\delta_{IJ} \delta_{KL} + g_{IJ}^{\phantom{IJ}P}g_{PKL} - f_{IJ}^{\phantom{IJ}P}f_{PKL} =
\frac{1}{N}\,\delta_{LI} \delta_{JK} + g_{LI}^{\phantom{LI}P}g_{PJK} - f_{LI}^{\phantom{LI}P}f_{PJK}\,. \, \, 
\ea
Therefore, the colored gravity provides a non-trivial Poisson algebra,
whose structure will become clearer
after a redefinition of generators
in the following section.

\subsection{Redefinition of Generators}

To make a contact with two-dimensional CFT dual, 
let us analyze the Poisson brackets obtained in the previous section
and reorganize the generators into the Virasoro primaries.
Firstly, the bracket $\{ \calL(\th_1)\,, \calL(\th_2) \}$ given in Eq.\eqref{PB1} 
is nothing but that of Virasoro algebra, and  
$\calL$ is its generator which creates boundary gravitons.
Secondly, the bracket $\{ \calJ^I(\th_1)\, , \calJ^J(\th_2) \}$ given in Eq.\eqref{PB2} satisfies the $SU(N)$ Kac-Moody algebra and suggests for us to interpret $\calJ^I$ as corresponding generators.
These are more or less expected results as $\calL$ and $\calJ^I$ are associated with the metric and the CS gauge connections, respectively.
We can make these statements more precise by considering 
the brackets among $\calL$, $\calJ^I$ and $\calK^I$.
The bracket \eqref{PB5} between $\calJ^I$ and $\calK^J$
shows that $\calK^J$ transforms in the adjoint representation of
 $SU(N)$ Kac-Moody algebra under the action of $\calJ^I$.
This is another hint of identifying $\calJ^I$ with $SU(N)$ Kac-Moody generators and does not raise any problem.

We have seen that the brackets \eqref{PB1}, \eqref{PB2} and \eqref{PB5} 
admit natural  CFT interpretations, 
but this is no longer true for  the brackets
\eqref{PB3} and \eqref{PB4}.
Firstly, the former bracket between
$\cJ^I$ and $\calL$  does not define Virasoro primaries for $\cJ^I$
as it simply commutes with  Virasoro.
Secondly,  in the latter bracket between $\cK^I$ and $\cJ^J$,
 the $\calK^I$ transforms almost like a primary of conformal dimension two but 
the transformation also involves
a nonlinear term proportional to $f^I{}_{JK}\,\cJ^J\,\cK^K$\,.
For the organization of the algebra as Virasoro primaries, we redefine the Virasoro generator as
\beq
\widetilde\calL (\theta) = \calL (\theta)  - \frac{2\pi}{\k}\, \calJ^I (\theta)\, \calJ_I (\theta)\,.
\eeq{Ltilde}
Then, its Poisson brackets with $\cJ^I$ and $\cK^I$ give the desired forms,
\bea
\{ \widetilde\calL (\theta_1) , \, \widetilde\calL(\theta_2) \, \}  \eq 
- \frac{N \k}{4 \pi} \,\partial^3 \d_{12}
-  \partial \widetilde\calL\,\d_{12} -2\,\widetilde\calL\,\partial \d_{12}  \, , \label{LLp} \\
\{ \calJ^I (\theta_1) , \widetilde\calL(\theta_2) \} \eq
-\partial\calJ^I  \d_{12} - \calJ^I  \partial \d_{12}\, ,  \label{LJp}\\
\{ \calK^I (\theta_1) , \widetilde\calL(\theta_2) \} \eq -\partial\calK^I \, \d_{12} - 2\,\calK^I \, \partial \d_{12}\,. 
\eea{LKp}
Therefore, in terms of new generators $\tilde\cL$\,,
the generators $\calJ^I$ and $\calK^I$  become Virasoro primaries of dimension one and two respectively.
As we do not modify $\cJ^I$ and $\cK^I$\,,
the generators $\calJ^I$ still define the $SU(N)$ Kac-Moody algebra, and 
the generators $\calK^I$ transform in the adjoint representation.

Finally, we are left to examine the bracket  $\{ \calK^I(\th_1) , \calK^J(\th_2) \}$  
given in Eq.\eqref{PB6}, which has non-linear terms.
Even taking the redefinition \eqref{Ltilde} into account,
it involves terms nonlinear in generators.
We find that this non-linearity is genuine in the sense that they cannot be removed by a redefinition of $\calK^I$ into
\be 
	\tilde\cK^I=\cK^I+\D\cK^I\,.
\ee
As the scaling dimension of $\calK^I$ is two, 
possible candidates for $\D\cK^I$ are 
 terms proportional to $\widetilde\calL$, $\partial\calJ^I$ and $\calJ^J\,\cJ^K$.
By asking that the $\tilde \cK^I$ --- hence the $\D\cK^I$ ---
transforms in the same way as the original $\cK^I$ under Virasoro and $SU(N)$ Kac-Moody algebras, 
we can immediately rule out  $\widetilde\calL$ and $\partial\calJ^I$ from the possibilities as they are not adjoint or primary.
Therefore, the only remaining possibility is  
\beq
\D\calK^I = {c^I}_{JK}\, \calJ^J \calJ^K\,, 
\eeq{Jquad}
where ${c^I}_{JK}$ is a constant symmetric under the exchange of the index $J$ and $K$\,. 
This modification does not spoil the fact that the $\tilde \cK^I$ is a Virasoro primary
of dimension two as
\beq
\{ \D\cK^I(\th_1),\widetilde\calL(\theta_2)  \} =
 -\partial\D\calK^I\, \d_{12} - 2\,\D\calK^I \, \partial \d_{12}\,.
\eeq{LJquad}
Meanwhile, $\D\calK^I$ transforms under the action of $\calJ^I$ as
\bea
\{ \D\cK^J(\theta_1), \calJ^I(\theta_2)  \} &=&  \frac{\k}{2 \pi} \,
c^{JI}{}_K\, \calJ^K\,\partial \d_{12}-2\,i
 {f^{IK}}_L\, {c^J}_{KM} \, \calJ^L\, \calJ^M\, \d_{12}\, .
\eea{JJquad}
Even though the second term can give
$-i\,f^{IJ}{}_K\,\D\cK^K$
by choosing $c^J{}_{KM}=g^J{}_{KM}$
and using the identity \eqref{g Jacobi},
the first term prevents $\D\cK^I$ from transforming
in adjoint representation.
With this, we exhausted all the possibilities of redefining $\calK^I$\,,
and the non-linearity of the bracket  $\{ \calK^I(\th_1) , \calK^J(\th_2) \}$ is 
free from ambiguities of generator redefinition.
Its form reads
\ba
&&\{ \cK^I (\theta_1), \cK^J (\theta_2) \} \, =-\, \frac{\k}{4\,\pi}\, \delta^{IJ}\, \partial^3\delta_{12} 
- \frac{\delta^{IJ}}{N} \( \partial\tilde\calL\, \delta_{12} + 2\,\tilde\calL \, \partial \delta_{12} \right)  \nn
&&\quad\ +\, i\, f^{IJ}_{\phantom{IJ}K}\left( \partial^2\cJ^K \, \delta_{12} +3\, \partial\cJ^K  \, \partial\delta_{12} +3\, \cJ^K \, \partial^2 \delta_{12} \right) 
- g^{IJ}_{\phantom{IJ}K}\left( \partial\cK^K \, \delta_{12} + 2\,\cK^K \, \partial \delta_{12} \right)\nn
&&\quad\ +\,i\frac{8\,\pi}{N\k}\, f^{IJ}_{\phantom{IJ}K}\,\tilde\calL \, \cJ^K  \, \delta_{12}+i\,\frac{4\,\pi}{\k}\left( f_{K}^{\phantom{K}IN} g_{LN}^{\phantom{LN}J} + g_{L}^{\phantom{L}IN} f_{KN}^{\phantom{KN}J} \right) \cJ^K \cK^L \, \delta_{12} \nn
&&\quad\ -\,\frac{4\,\pi}{\k}\,
\left(\frac{\d^{IJ}\,\delta_{KL}}{3\,N}-f_{K}^{\phantom{K}IN} f_{LN}^{\phantom{LN}J}\) \left(\partial\cJ^K \, \cJ^L \,  \delta_{12} +2\,\cJ^K \, \partial \cJ^L  \, \delta_{12}+3\,\cJ^K  \,\cJ^L \, \partial \delta_{12} \right) \nn
&&\quad\ +\,i \left(\frac{4\,\pi}{\k} \right)^2 
\( \frac{\d_{KL}}{N}\,{f_M{}^{IJ}} - f_{K}^{\phantom{K}IO} f_{LO}^{\phantom{LO}P} f_{MP}^{\phantom{MP}J}\) \cJ^K \cJ^L \cJ^M \delta_{12}\,. \label{PB6_prime}
\ea  
The result of this Poisson bracket involves essentially all allowed terms of dimension three.
The central term is proportional to $\partial^3\d$.
The linear part involves contributions from all the three generators $\cL$, $\cJ^I$ 
and $\cK^I$ properly adjusted with derivatives.
The quadratic part contains $\tilde\cL\,\cJ^I$, $\cJ^I\,\cK^J$ and $\cJ^I\,\cJ^J$ terms. Finally, the cubic part only involves $\cJ^I\,\cJ^J\,\cJ^K$ without any derivatives.
{See \cite{Sug} and \cite{Henneaux:2012ny} for a similar type of nonlinearities appearing in extended AdS supergravities and in extended higher-spin AdS supergravities, respectively.}

\subsection{Mode Expansion}
For the completeness, we also provide the Poisson bracket in terms of Fourier modes:
\be
	\cL_n=-\cL[e^{i\,n\,\theta}]-\frac{N\,\k}4\,\delta_{n,0}\,,
	\qquad
	\cJ^I_n=i\,\cJ^I[e^{i\,n\,\theta}]\,,
	\qquad
	\cK^I_n=\cK^I[e^{i\,n\,\theta}]\,,
\ee
where we replaced the notation for $\tilde\cL$ to $\cL$
and introduced factors to get more standard expressions.
Recall that the $SU(N)$ CS connection is anti-Hermitian \eqref{reality}
where the Hermitian generators $\bm T^I$ is used for $\mathfrak{su}(N)$\, algebra. As a result, the generator $\cJ^I(\theta)$ is purely imaginary.
With the $i$ factor, the generators now satisfy
\be
	\cL_n^\dagger=\cL_{-n}\,,
	\qquad 
	{\cJ^I_n}^\dagger=\cJ^I_{-n}\,,
	\qquad
	{\cK^I_n}^\dagger=\cK^I_{-n}\,.
\ee
First, the brackets containing Virasoro generators are
\ba
i\, \{ \calL_n , \calL_m \} \eq \frac{N\k}{2}\, (n^3-n)\, \delta_{n+m,0} + (n-m)\,\calL_{n+m}\,, 
\label{LL com} \\
i \,\{ \calL_n , \cJ^I_m \} \eq - m\,\cJ^I_{n+m}\,, 
\label{LJ com}\\
i\, \{ \calL_n , \cK^I_m \}\eq (n-m)\,\cK^I_{n+m}\,.
\label{LK com}
\ea
The other brackets containing $\cJ^I_n$ give
\ba
i\, \{ \cJ^I_n , \cJ^J_m \} \eq
 -\frac{\k}{2}\,\d^{IJ}\,n \, \delta_{n+m,0} - i\, f_{K}^{\phantom{K}IJ} \cJ^K_{n+m}\,,\\
i\, \{ \cJ^I_n , \cK^J_m \} \eq - i\, f_{K}^{\phantom{K}IJ} \cK^K_{n+m}\,,
\label{JK com}
\ea
Finally, the bracket between two $\cK^I$s reads
\ba
i\, \{ \cK^I_n , \cK^J_m \} \eq
 \frac{\k}{2}\,\delta^{IJ}\,(n^3-n)\, \delta_{n+m,0} +
\frac{\delta^{IJ}}{N}\, (n-m)\,\calL_{n+m}  \nonumber \\
&&+i\,(n^2 -n\,m +m^2-1)\,f_{K}^{\phantom{K}IJ}\,\cJ^K_{n+m} -(n-m)\,g_{K}^{\phantom{K}IJ}\cK^K_{n+m} \nonumber \\
&&+i \,\frac{4}{N\k}\,f_{K}^{\phantom{K}IJ} \sum_l \calL_{n-l}\,\cJ^K_{m+l}
-i\frac{2}{\k}
\left( f_{K}^{\phantom{K}IN} g_{LN}^{\phantom{LN}J} + g_{L}^{\phantom{L}IN} f_{KN}^{\phantom{KN}J} \right)
 \sum_l \cJ^K_{n-l}\, \cK^L_{m+l} \nonumber \\
&&+\, \frac{2}{\k} 
\left(\frac{\d^{IJ}\,\delta_{KL}}{3\,N}-f_{K}^{\phantom{K}IN} f_{LN}^{\phantom{LN}J}\)
 \sum_l (n-m+l)\, \cJ^K_{n-l}\, \cJ^L_{m+l} \nonumber \\
&&+i\( \frac{2}{\k} \)^2
\( \frac{\d_{KL}}{N}\,{f_M{}^{IJ}} - f_{K}^{\phantom{K}IO} f_{LO}^{\phantom{LO}P} f_{MP}^{\phantom{MP}J}\)
\sum_{k,l} \cJ^K_{k}\, \cJ^L_{l}\, \cJ^M_{n+m-k-l}\,.
\label{pb last}
\ea
Comparing Eqs.(\ref{LL com}\,--\,\ref{LK com}) with the standard definition of  Virasoro and Kac-Moody subalgebras,
\ba
	\[ L_n,L_m\] \eq \frac{c}{12}\,(n^3-n)\,\d_{n+m,0}+(n-m)\,L_{n+m}\,,\\
	\[ L_n, J^a_m\] \eq -m\,J^a_{n+m}\,,\\
	\[ J^a_n,J^b_m\] \eq k_{\rm\sst KM}\,\delta^{ab}\,n\,\d_{n+m,0}+i\,f^{ab}{}_c\,J^c_{n+m}\,,
	\label{Gen c k}
\ea
we find that the Virasoro subalgebra generated by $\cL_n$ 
has the central charge,
\be
	c=6\,N\,\k=\frac{3\,\ell}{2\,G}\,. 
	\label{c Nk}
\ee
This coincides with the central charge of the pure Einstein gravity.
Turning to the Kac-Moody subalgebra generated by $\cJ^I_n$\,,
we find that the central term has the coefficient,
\be
	k_{\rm\sst KM}=-\frac{\k}2\,.
\ee
When the color gauge group $SU(N)$ is fixed,
the ratio between these coefficients is fixed by $c/k_{\rm\sst KM}=-12\,N$\,.
Note that the 
central term of the Kac-Moody algebra
comes with negative sign. 
We shall come back to this important point later.

\subsection{Rigidity of the structure}
\label{sec: rigidity}

A natural question at this point is whether the Poisson algebra
we have obtained can be generalized such that
the two central terms $c$ and $k_{\rm\sst KM}$ become independent of each other.
In order to examine this idea, we begin with an ansatz,
\ba
i\, \{ \cK^I_n , \cK^J_m \} \eq
C\,\delta^{IJ}\,(n^3-n)\, \delta_{n+m,0} +
C_{\cL}\, (n-m)\,\calL_{n+m}  \nonumber \\
&&+\,C_{\cJ}\,(n^2 -n\,m +m^2-1)\,f_{K}^{\phantom{K}IJ}\,\cJ^K_{n+m} +
C_{\cK}\,
(n-m)\,g_{K}^{\phantom{K}IJ}\cK^K_{n+m} \nonumber \\
&&+\,C_{\cL\cJ}\,f_{K}^{\phantom{K}IJ} \sum_l \calL_{n-l}\,\cJ^K_{m+l}
+C_{\cJ\cK}
\left( f_{K}^{\phantom{K}IN} g_{LN}^{\phantom{LN}J} + g_{L}^{\phantom{L}IN} f_{KN}^{\phantom{KN}J} \right)
 \sum_l \cJ^K_{n-l}\, \cK^L_{m+l} \nonumber \\
&&+\,C_{\cJ\cJ}
\left(\frac{\d^{IJ}\,\delta_{KL}}{3\,N}-f_{K}^{\phantom{K}IN} f_{LN}^{\phantom{LN}J}\)
 \sum_l (n-m+l)\, \cJ^K_{n-l}\, \cJ^L_{m+l} \nonumber \\
&&+\,C_{\cJ\cJ\cJ}
\( \frac{\d_{KL}}{N}\,{f_M{}^{IJ}} - f_{K}^{\phantom{K}IO} f_{LO}^{\phantom{LO}P} f_{MP}^{\phantom{MP}J}\)
\sum_{k,l} \cJ^P_{k}\, \cJ^Q_{l}\, \cJ^R_{n+m-k-l}\,,
\ea
with eight arbitrary constants,
\be
	C\,,\quad C_{\cL}\,,\quad C_{\cJ}\,,\quad C_{\cK}\,,
	\quad C_{\cL\cJ}\,,\quad C_{\cJ\cK}\,\quad C_{\cJ\cJ}\,,
	\quad C_{\cJ\cJ\cJ}\,.
	\label{Cs}
\ee
For the other Poisson brackets, we take 
Eqs.\eqref{LK com}, \eqref{JK com}, 
the classical counterpart of Eqs.\eqref{Gen c k}.
We examine Jacobi identity on this ansatz to find out all consistent 
sets of $(c=6\,N\,\k,k_{\rm\sst KM})$ and the  constants in Eq.\eqref{Cs}.
The Jacobi identity between one $\cL$ and two $\cK$ gives the relations,
\begin{align}
C = \frac{N\k}{2}\, C_{\cL}\,, \qquad C_{\cJ} = \frac{N\k}{4}\,C_{\cL \cJ}\,,
\label{JR1}
\end{align} 
whereas the Jacobi identity between $\cJ$ and two $\cK$ gives
\begin{align}
C = i \,k_{\rm\sst KM}\, C_{\cJ} \,, \quad C_{\cJ} = -i \,k_{\rm\sst KM}\, C_{\cJ \cJ} \,, \quad C_{\cK} = i \,k_{\rm\sst KM}\, C_{\cJ \cK} \,, \quad C_{\cJ \cJ} = -i\, k_{\rm\sst KM}\, C_{\cJ \cJ \cJ} \,.
\label{JR2}
\end{align}
The six linear relations in Eq.\eqref{JR1} and Eq.\eqref{JR2} determine the constants
 in terms of $k_{\rm\sst KM}$, $C$ and $C_{\cK}$. 
 Finally, the Jacobi identity between three $\cK$'s gives
\begin{align}
k_{\rm\sst KM} = -\frac{\k}{2} \,, \quad C = \frac{\k}{2} \, C_{\cK}{}^2 \,.
\end{align}
In this way, all the constants are fixed in terms of $C_{\cK}$\,,
and the latter is not restricted by Jacobi identities.
Actually, the arbitrary constant  $C_{\cK}$ 
can be absorbed into a redefinition of $\cK$'s.
Therefore, once we fix the normalization of $\cK$, then 
there is no freedom left.

\section{Discussions}\label{sec: 5}

We can summarize what we have achieved in this paper as follows.
Starting from  the $SU(N,N)_L\times SU(N,N)_R$
CS theory which admits the interpretation of colored gravity with the diffeomorphism adjoined by the color group $SU(N)$\,, we have obtained a non-linear Poisson algebra
generated by spin-2 generators $\cL_n$, $\cK^I_n$ and spin-1 generators $\cJ^I_n$ as the asymptotic symmetry algebra of the theory.
The generators $\cL_n$ and $\cJ^I_n$ form the standard Virasoro and Kac-Moody subalgebras, whereas  the brackets of two $\cK^I_n$s,
which are Virasoro primary of dimension-two and adjoint under Kac-Moody symmetry, generate non-linear structures. 

The algebra we obtained is new and worth exploring further. An outstanding question is to identify a class of two-dimensional `flavored' conformal field theories whose holographic dual correspond to the colored gravity at various vacua. At finite temperature, these conformal field theories would be described by some colored counterpart of the BTZ black holes. An extension of the analysis put forward in \cite{Rey:2006bz} should reveal the new correspondence more concretely. {A related question is to understand how breaking and restoration of $(1+1)$-dimensional Lorentz symmetry is reflected in the (non)-unitarity of the two-dimensional CFT. Such issues were discussed already in other contexts \cite{Gary:2014mca, Lee:2017utr}. }

In the rest of this section, we devote to discuss aspects of the algebraic structure and also supersymmetric extensions relevant for colored counterpart of three-dimensional AdS supergravity.  
{Let us also mention
that analogous issues  in the higher-spin context  and in a wider class of examples 
have been discussed in \cite{Castro} and \cite{Afshar}, respectively.}

\subsection{Non-unitarity of Kac-Moody Algebra Sector}

An intriguing point of the asymptotic algebra we obtained above is 
that the central term for the $SU(N)$ Kac-Moody subalgebra takes a negative value. This can also be checked by examining the Jacobi identities for a generic set of coefficients with the ansatz \eqref{Cs} in Section \ref{sec: rigidity}. For the Kac-Moody algebra alone, the negative central term does not make any problem since we can simply redefine the 
generator $\cJ^I_n$ into $\cJ^I_{-n}$, which is equivalent to interchanging the 
left-moving and right-moving sectors. However, in the current case, we cannot utilize this redefinition as it will violate the natural grading of the algebra, $[\cA_m, \cA_n]\subset \cA_{m+n}$\,, where $\cA_n$ is the subspace of  polynomials in $\cL_n$, $\cJ^I_n$, $\cK^I_n$ with $\cL_0$ eigenvalue $n$\,.
Hence, unless the Kac-Moody central term changes its sign upon quantization of the algebra for other reasons, the usual lowest-weight representations would involve negative norm states created by $\cJ^I_{-n}$'s.

We claim that this non-unitarity is a distinguishing feature of the colored gravity in three dimensions. Recall that we already encountered non-unitarity when we consider the linearized spectrum around a color-symmetry breaking rainbow vacuum. When a background solution is chosen such that the color-symmetry is broken down as Eq.\eqref{broken}, we have 
\begin{itemize}
\item $(N-k)^2$ unitary and $k^2$ non-unitary spin-two fields, 
\item $k^2$ unitary and $(N-k)^2-1$ non-unitary spin-one fields, and
\item $2\,(N-k)\,k$ partially-massless spin-two fields.
\end{itemize}
Let us recall that a left-moving partially-massless spin-two has two boundary degrees of freedom of which only one is unitary.
Thus, if we only collect the unitary modes,
then there are 
\begin{itemize}
\item $(N-k)^2$ spin-two fields, 
\item $k^2$ spin-one fields, and
\item  $2\,(N-k)\,k$ partially-massless spin-two fields.
\end{itemize}
In total, there are exactly $N^2$ unitary modes.
On the other hand, a similar counting shows that
there are $(N^2-1)$ non-unitary modes:
\begin{itemize}
\item $k^2$ spin-two fields, 
\item $(N-k)^2-1$ spin-one fields, and
\item  $2\,(N-k)\,k$ partially-massless spin-two fields.
\end{itemize}
The partially-massless spin-two fields are doubly counted since they have one unitary and one non-unitary mode in left (or right)-moving sector.
It is important to note that the respective numbers of unitary modes and non-unitary modes do not depend on the value of $k$\,. Hence, as $k$ varies, that is to say, as the vacuum changes from one color-symmetry breaking pattern to another pattern, the number of positive and negative norm states do not change and just rearrange into spin-two, spin-one and partially-massless spin-two representations. This suggests that the non-unitarity is stringently tied up  with the underlying non-compact group structure (which may eventually extend to supergroup structure) when graviton is color-decorated. 

To be more precise, let us observe that the Poisson bracket between dynamical variables $\varphi_i ( x)$ and $\varphi_j (y)$ takes the form of
\beq
\{ \varphi_i ( x) , \varphi_j (y) \} \sim \kappa\, d_{ij}\, \delta (x - y) + \cdots , 
\eeq{structure_PB}
where $\kappa$ is the CS level and $d_{ij}$ is the  Killing inner product of $\mathfrak{su}(N,N)$ between generators associated with $\varphi_i$ and $\varphi_j$. 
In the basis \eqref{generalform}, it is clear that the $2\,N^2$ generators of spin-two fields --- $J_1\otimes \bm I, J_1\otimes \bm T^I, J_2\otimes \bm I$ and $J_2\otimes \bm T^I$ --- have positive Killing inner products while the other $N^2$ generators --- $J_0\otimes \bm I$ and $J_0\otimes \bm T^I$ --- have negative ones.
But the asymptotic behavior of fields reduces the number of physical fields to $N^2$, which were previously identified with one color-singlet spin-two field $\mathcal{L}$ and $(N^2 -1)$ colored spin-two fields $\mathcal{K}^I$.
The fact that all the spin-two fields generate positive norm states accounts for the positive central terms in the Possion brackets $\{ \calL(\theta_1), \calL(\theta_2) \}$ and $\{ {\cal K}^I(\theta_1) , {\calK}^J(\theta_2) \}$.
On the other hand, it is also clear that the $(N^2 -1)$ generators of spin-one fields ---  $I\otimes \bm T^I$ --- have negative Killing inner products.
Since spin-one fields get no further constraints from the asymptotic behavior, all these fields appear in the asymptotic algebra in terms of ${\calJ}^I$ to give negative level of the Kac-Moody algebra.
Different vacua will mix up these physical states among themselves to give a different spectrum, yet they will retain the number of unitary modes and non-unitary modes as $N^2$ and $(N^2 -1)$ at any of the $N$ vacua.

\subsection{Colored Higher Spin Gravity}

The analysis of the current paper can be extended to various direction.
One particularly interesting and important direction is the CS colored higher-spin gravity with the gauge algebra,
\be
	\mathfrak{g}=hs(\l)\otimes \mathfrak{u}(N) \ominus {\rm id}\otimes \mathbb I\,.
\ee
In fact, when $\lambda=1/2$ and $N=2^{[\cN/2]}$\,, the above algebra coincides with the bosonic 
sector of supersymmetric higher-spin algebra $shs(\cN|2,\mathbb R)$ \cite{Henneaux:2012ny} which 
is the higher-spin extension of the gravity sector $\mathfrak{osp}(\cN|2,\mathbb R)$
generated by 
\be
	J_a=q_\a\,\,q_\b\,\s_a^{\a\b}\,,
	\qquad Q^i_\a=q_\a\,\psi^i\,,
	\qquad R^{ij}=\psi^i\,\psi^j\,.
\ee
The oscillators $q_\a$ and $\psi^i$ are respectively Grasmannian even and odd
and the product between generators are realized by the usual star product of the oscillators.
The higher spin version contains generators of type:
\be
	M_{\a_1\cdots \a_n}{}^{i_1\cdots i_m}
	=q_{\a_1}\cdots q_{\a_n}\,\psi^{i_1}\cdots \psi^{i_m}
	\qquad [\,n+m\in 2\,\mathbb{Z}\,]\,,
\ee
and the spin of a generator is given by $s=n/2$\,.
Focusing on a particular bosonic spin $s$\,, there are multiple of generators
whose Grasmannian odd part
 generated by
\be
	\psi^{i_1}\cdots \psi^{i_{2p}} \qquad \big[\,p=0,1,\ldots, [\,\tfrac{\cN}2]\,\big]\,.
\ee
The polynomials of $\psi^i$ form a Clifford even algebra, isomorphic to $\mathfrak{u}(2^{[\frac{\cN}2]})$\,.
In \cite{Henneaux:2012ny}, the asymptotic symmetry of the supersymmetric CS higher spin 
gravity with the algebra $shs(\cN|2,\mathbb{R})$ 
has been analyzed in details and further classified all other possible types of the supersymmetric CS higher spin gravity. 
{In \cite{Candu}, the asymptotic symmetry of the matrix extended $shs(\cN|2,\mathbb{R})$  higher spin gravity has been studied. 
However, in this work, the Poisson brackets between colored generators of spin $s\ge2$ were not explicitly analyzed.} 
Therefore, it would be worth to revisit the analysis focusing now on the `color' aspects. We would not be surprised to discover surprising richness both in algebraic structure and in physics implications.

\acknowledgments
We are grateful to Seungho Gwak, Marc Henneaux, Daniel Grumiller, Karapet Mkrtchyan, Alexander Polyakov and Massimo Porrati for numerous useful discussions. We also thank hospitality of CCPP of New York University, Physics Department of Princeton University, Erwin Schr\"odinger Institute of Mathematical Physics and Theoretical Physics group of Technische Universit\"at Wien, where part of this work was carried out. This work was supported in part by the National Research Foundation (Korea) through the grant 2014R1A6A3A04056670 (EJ), 
and the grants 2005-0093843, 2010-220-C00003 and 2012K2A1A9055280 (JWK, SJR). 

\appendix

\section{Colored gravity basis of $\mathfrak{su}(N,N)$}\label{app}

Let us remind that we are working with  CS theory whose gauge algebra  is $\mathfrak{g} \oplus \mathfrak{g}$, where the algebra $\mathfrak{g}$ is isomorphic to $\mathfrak{su}(N, N)$.
An element $X$ of $\mathfrak{su}(N, N)$ is a traceless $2N\times 2N$ complex matrix
with the condition,
\be
	X^\dagger \(\begin{array}{cc} \bm I& \bm 0 \\ \bm 0& -\bm I \end{array} \)+\(\begin{array}{cc} \bm I& \bm 0 \\ \bm 0& -\bm I \end{array} \) X=0\,.
\ee
Here, $\bm{I}$ is the $N \times N$ identity matrix.
The isomorphism $\mathfrak{g}\simeq \mathfrak{su}(N,N)$ can be seen by expressing $X$ in the block matrix form,
\beq
X = \(\begin{array}{cc} \bm A& \bm B \\ \bm B^{\dagger}& \bm C \end{array} \),
\eeq{block}
where $\bm A, \bm B$ and $\bm C$ are $N \times N$ matrices satisfying $\bm A^\dagger=-\bm A$, 
$\bm C^\dagger=-\bm C$ and $\mathrm{Tr}(\bm A+\bm C) =0$.
We can expand these matrices in terms of $\bm I$ and 
the Hermitian generators $\bm{T_I}$ of $\mathfrak{su}(N)$ to have
\be
\bm A = i\,(a^0\,\bm{I} + a^I\,\bm{T_I})\,, \qquad 
\bm B = b^0\,\bm{I}+ b^I\, \bm{T_I}  ~, \qquad  
\bm C = i\,c^0\, \bm{I} + i\,c^I\, \bm{T_I}\,.
\label{expand}
\ee
Here the coefficients  $a^0, a^I$, $c^I$ and $c^0$ are real with $a_0+c_0=0$ while 
$b^0$ and $b^I$ are complex numbers.
Redefining these coefficients as
\be
a^0=x^0\,, \quad a^I=y^I+z^{0I}\,,\quad
b^0=x^1+i\,x^2\,, \quad b^I=z^{1I}+i\,z^{2I}\,,\quad
c^I=y^I-z^{0I}\,,
\label{new_coeff}
\ee
we can express an arbitrary element $X$ of $\mathfrak{su}(N, N)$  as
\beq
	X =
	x^a \,J_a \otimes \bm{I} + y^I\,i\,I \otimes \bm{T_I} +  z^{aI}\,J_a\otimes \bm{T_I}\,,
\eeq{generalform}
making use of the $\mathfrak{u}(1,1)$ generators,
\beq
I= \( \begin{array}{cc}  1 & 0 \\ 0 & 1 \end{array} \) , \qquad
J_0 = \( \begin{array}{cc}  i & 0 \\ 0 & -i \end{array} \) , \qquad 
J_1 = \( \begin{array}{cc}  0 & 1 \\ 1 &  0 \end{array} \) , \qquad 
J_2 = \( \begin{array}{cc}  0 &  i\\  -i & 0  \end{array} \).
\eeq{su11}
The isomorphism between $\mathfrak{u}(1,1)$ and $\mathfrak{gl}(2, \mathbb{R})$ \eqref{gl2} now shows that 
the underlying algebra of the colored gravity, presented in the basis \eqref{generalform},  is 
indeed isomorphic to $\mathfrak{su}(N,N)$.


\end{document}